# Effect of Doping and In-Composition on Gain of Long Wavelength III-Nitride QDs

Ahmed S. Jbara[1] and H. I. Abood[2], Amin H. Al-Khursan[3,*]

[1]Physics Department, Science College, Al-Muthana University, Samawah, Iraq.

[2]Physics Department, Science College, Babylon University, Hillah, Iraq.

[3]Nassiriya Nanotechnology Research Laboratory (NNRL), Science College, Thi-Qar University, Nassiriya, Iraq.

* Corresponding author: ameen_2all@yahoo.com, (Mobile: 00964 7813809600)

**Abstract:** In this work, we calculate material gain for long wavelength III-nitride InN and AlInN quantum dot (QD) structures. Strain and QD inhomogeneity are included in the calculations. The study covers (800-2300 nm) wavelength range which is important in optical communications. While p-doping is shown to be efficient to increasing gain, changing QD size (especially QD radius) is more efficient to vary wavelength. The results predicted that n-doped QD structures are promises for broad band laser applications.

## 1. Introduction

As a new materials system, group III- nitride (III-N) based devices are of particular interest due to their wide range of emission frequencies and their potential for high-power electronic applications [1]. Among the group III-N semiconductors, InN is a key material for optical and high-temperature device applications. It is reported that InN with a bandgap energy (0.64 eV) [2] shows unusual physical properties, anisotropic critical field [3], unique transport and optical properties [4]. This results from the large built-in electric field which arises from strain associated with these lattice mismatched hexagonal (wurtzite- W) III-nitride structures. In recent years, nitride-based quantum dots (QDs) has been the subject of intense experimental and theoretical research due to their promising applications in lasers and optical amplifiers as well as possible applications for memory storage and quantum computing. Its importance also results from the possibility to engineering it simply by changing the dot size or composition to emit anywhere from the infrared to ultraviolet [5].

It is important to use a simple technique to calculate electronic and optical properties of these nitride-based QDs. Thus, we study here the optical properties of some nitride QD structures beginning from the calculation of their electronic properties where the strain is taken into account. QD size and shape fluctuations are included by assuming that gain is inhomogeneously broadened. Both QD size and composition are addressed. WL composition is also addressed. Accordingly, our study covers the range (800-2300 nm). We include all the wetting layer (WL) states in the calculation of quasi-Fermi energy. Both types of doping are also examined.

## 2. Theory

The material gain per QD layer of a self-assembled QD is expressed as [6]

$$g(\hbar w)=\frac{\pi e^2}{n_b c \varepsilon_0 m_0^2 w}\sum_i \int_{-\infty}^{\infty} dE' \left|M_{env}\right|^2 \left|\hat{e}.P_{cv}\right|^2 D(E')L_g(E',\hbar w)\left[f_c(E',F_c)-f_v(E',F_v)\right] \quad (1)$$

where the summation over $i$ is carried out to account for all radiative transitions, $e$ is the electronic charge, $n_b$ is the background refractive index of the material, $c$ is the speed of light in free space, $\varepsilon_o$ is the permittivity of free space, $m_o$ is the free electron mass, $w$ is the angular optical frequency, and $E'$ is the optical transition energy. The term $\left|M_{env}\right|^2$ is the envelope function which overlaps between the QD electron and hole states. The term $\left|\hat{e}.P_{cv}\right|^2$ is the momentum matrix of the QD depending on the polarization of light under the parabolic band model. $\hat{e}$

is a unit vector in the polarization direction. The momentum matrices of QDs are taken to be the same as those of QW near the zone center ($k_t = 0$) and are expressed as $|\hat{e}.P_{cv}|^2 = 3/2(M_b^2)$ where the bulk momentum matrix element $M_b^2 = (m_0/6)E_p$, where $E_p$ is the optical matrix energy parameter for the W crystals, the term $D(E')$ represents the inhomogeneously broadened density of states for self-assembled QDs. When the spectral variance of QDs is $\sigma$ and the transition energy at the QD maximum distribution of the $i^{\text{th}}$ optical transition is $E_{mac}^i$, $D(E')$ is given by

$$D(E') = \frac{s^i}{V_{dot}^{eff}} \frac{1}{\sqrt{2\pi\sigma^2}} \exp\left( \frac{-(E' - E_{\max}^i)^2}{2\sigma^2} \right) \qquad (2)$$

here $s^i$ represents the degeneracy number at each QD state. The term $V_{dot}^{eff}$ represents the effective volume of QDs, it is expressed as $V_{dot}^{eff} = h/N_D$, where $h$ is the average height of QDs and $N_D$ is the areal density of QDs. $L_g(E', \hbar w)$ is the Gaussian lineshape function, it is expressed as

$$L_g(E', \hbar w) = \frac{1}{\sqrt{2\pi\gamma^2}} \exp\left( \frac{-(E' - \hbar w)^2}{2\gamma^2} \right) \qquad (3)$$

where $\gamma$ is the variance of the linewidth. The terms $f_c$ and $f_v$ represent the respective quasi-Fermi distribution function for the conduction and valence bands [6].

## 3. Calculated Results

We study two types of structures, they are: $InN/In_xAl_{1-x}N/In_{0.25}Al_{0.75}N$ for x=0.76-0.96 and $In_xAl_{1-x}N/In_{0.86}Al_{0.14}N/In_{0.25}Al_{0.75}N$ for x=0.88-0.96. Thus, the effect of changing both QD and WL composition is examined to obtain III-N QDs emitting at long wavelength region. Because of its importance, strain is taken into account in the structures studied. The energy gap for $In_xAl_{1-x}N$ alloy can be calculated from the relation [7]

$$E_g(x) = x.E_{g\,InN} + (1-x).E_{g\,AlN} - x(1-x).b_{In} \qquad (4)$$

where $E_{g\,InN}$, $E_{g\,AlN}$ are the energy bandgaps for InN and AlN, respectively. $b_{In}$ is the bowing parameter related to In-composition. Commonly used values for $b_{In}$ in hexagonal III-N alloys are listed in Table (1). Accordingly, the bandgaps of QD and WL are states in Table (2). Spontaneous polarization is not taken into account here. Energy subbands for these structures are calculated using quantum disc model [5]. Gain calculations are done at four values of the surface electron densities per QD layer ($n_{2D}$), they are: $5.2\times10^{14}$, $6.2\times10^{14}$, $7.3\times10^{14}$ and $8.3\times10^{14}$ $m^{-2}$, the areal density of QDs is $1\times10^{14}$ $m^{-2}$. The effect of indium composition in the QD and WL and the quantum-size effects are investigated for the undoped, p-doped and n-doped structures where the surface carrier density of ($3\times10^{15}$ $m^{-2}$) of p- or n-type ionized acceptors per QD layer are used. Finally, we compare the p- and n-doped effects.

### 3.1 p-Doping Effect

An example of gain spectra appears in Fig. 1 which shows gain curves for InN QDs with and without p-doping while Fig. 2 shows the effect of doping on InAlN QDs. At the same carrier density, p-doping is shown to increases material gain of InN (InAlN) QDs by ~ 3 (3.5) times, it blue-shifts the wavelength by ~ 20 (30) nm. Thus, p-doping has an efficient effect on III-N QDs. It is known [8] that p-doping increases gain. Hole states are more than electron states due to their larger effective masses then, they closely spaced. So, holes are thermally broadened between excited states instead of ground state. Thus, p-doing provides an excess hole concentration in the GS which results in increasing gain.

### 3.2 n-Doping Effect

Doping with n-type dopants is also addressed. Fig. 3 shows the effect of n-doping on InN QDs, where it is shown to increase gain and blue-shifts wavelength. A comparison between undoped, p- and n-doped spectra is shown in Fig. 4 for InN QDs and Fig. 5 for InAlN QD structure. While p-doping is shown to increase gain to higher value, a few increment in gain is shown for n-doping. This is because while p-doping increases holes in the valence band, which have a small concentration before doping, here n-doping increases electron concentration which is enough before n-doping. So gain increment is not too much. An interesting feature can be seen with n-doping. The gain bandwidth for both n-doped structures increases by 150 (100) nm for InN (InAlN) QDs. This promises to use n-doped QD structures in broad-band lasers applications.

From these figures InN QDs have the highest gain obtained while InAlN QDs shows higher increment ratio for both p- and n-doping. The ratio of (p-doped/undoped) for InN gain increases by 3 times, while for InAlN it increases by 3.5 times. For (n-doped/undoped) the increment ratio is 1.4 for InN and 1.6 for InAlN.

### 3.3 Composition Effects

Changing QD and WL composition is also addressed. Fig. 6 shows the effect of changing WL composition on InN QD structures. Gain increases and red-shifts with increasing In-composition in the WL till x=0.88. For higher composition ratio, gain is reduced. Fig. 7 shows the effect of changing QD composition where the gain increases and red-shifts with increasing In-composition as expected in [9].

### 3.4 QD Size Effect

It is expected that reducing dot size gives higher gain. The quantum size effects (disc height and radius) in both gain and wavelength is shown in Fig. 8 (a), (b) and Fig. 9 (a), (b), respectively for the two types of structures studied.

Fig. 8 (a), (b) shows the peak material gain as a function of the dick height and radius, respectively. Reducing disc height is more efficient than radius of the dot for gain increment. Both undoped and p-doped structures show this increment. For dot height and radius reduction, the increment ratio of (doped/undoped) QD gain is approximately the same for both InN and AlInN QD structures. Doping InN QDs is shown to be more efficient to gives higher gain, the highest gain increment ratio is shown for doped InN QDs (800 $cm^{-1}$/nm).

Long wavelength emission from nitride structures is an important task in recent optical communication researches. So we study it in Fig. 9 (a), (b) which shows the peak wavelength as a function of the dick height and radius. Wavelength is red-shifted with increasing disc size (height or radius) while it blue-shifted with doping. Changing size is more efficient than doping to vary wavelength. While wavelength is approximately still constant when smaller QDs are doped for both structures, high shift (~100 nm) is obtained when the InN QD radius changes from 13 to 15 nm.

## 4. Conclusions and Future Development

Depending on the above results, linear gain is calculated taking QD inhomogeneity into account. Through the calculations, all WL subbands are included in the Fermi-energy calculations. The following findings can be stated: The effect of p-doping is efficient on long wavelength III-N QDs by increasing gain and blue-shifts wavelength, n-doping has less effect, for higher In-composition in the WL gain is reduced and red-shifted, while gain is increased and red-shifted with increasing In-composition in the QD, and gain increases with reducing QD size and wavelength is blue-shifted.

Other properties for these structures need to be examined like: The gain relation stated in this work can be used to calculate lattice temperature effects, and the effect of p-doping on radiative and nonradiative characteristics of III-N QDs.

## Acknowledgment

We acknowledge our indebtedness to Prof. Mohammed Jasim Betti (Dept. of English, College of Education, Thi-Qar University) for proofreading the language of this work.

## Tables

Table (1): Calculated bandgaps of the QD, WL and barrier layers of structures studied. (a) when WL is changed, (b) when QD layer is changed.

**Table (1a)**

| *InN / $In_xAl_{1-x}N$ / $In_{0.25}Al_{0.75}N$* | | | |
|---|---|---|---|
| ***Properties*** | ***Quantum Dot*** | ***Wetting Layer*** | ***Barrier*** |
| ***Material*** | *InN* | $In_{0.96}Al_{0.04}N$ | $In_{0.25}Al_{0.75}N$ |
| ***Eg*** | **0.64** | **0.739** | **4.097** |
| ***Material*** | *InN* | $In_{0.92}Al_{0.08}N$ | $In_{0.25}Al_{0.75}N$ |
| ***Eg*** | **0.64** | **0.848** | **4.097** |
| ***Material*** | *InN* | $In_{0.88}Al_{0.12}N$ | $In_{0.25}Al_{0.75}N$ |
| ***Eg*** | **0.64** | **0.966** | **4.097** |
| ***Material*** | *InN* | $In_{0.84}Al_{0.16}N$ | $In_{0.25}Al_{0.75}N$ |
| ***Eg*** | **0.64** | **1.094** | **4.097** |
| ***Material*** | *InN* | $In_{0.80}Al_{0.20}N$ | $In_{0.25}Al_{0.75}N$ |
| ***Eg*** | **0.64** | **1.232** | **4.097** |
| ***Material*** | *InN* | $In_{0.76}Al_{0.24}N$ | $In_{0.25}Al_{0.75}N$ |
| ***Eg*** | **0.64** | **1.379** | **4.097** |

**Table (1b)**

| $In_x.Al_{1-x}N/In_{0.86}Al_{0.14}N$ / $In_{0.25}Al_{0.75}N$ | | | |
|---|---|---|---|
| ***Properties*** | ***Quantum Dot*** | ***Wetting Layer*** | ***Barrier*** |
| ***Material*** | $In_{088}.Al_{0.12}N$ | $In_{0.86}Al_{0.14}N$ | $In_{0.25}Al_{0.75}N$ |
| ***Eg*** | **0.966** | **1.029** | **4.097** |
| ***Material*** | $In_{0.9}Al_{0.1}N$ | $In_{0.86}Al_{0.14}N$ | $In_{0.25}Al_{0.75}N$ |
| ***Eg*** | **0.906** | **1.029** | **4.097** |
| ***Material*** | $In_{0.92}Al_{0.08}N$ | $In_{0.86}Al_{0.14}N$ | $In_{0.25}Al_{0.75}N$ |
| ***Eg*** | **0.848** | **1.029** | **4.097** |
| ***Material*** | $In_{0.94}Al_{0.06}N$ | $In_{0.86}Al_{0.14}N$ | $In_{0.25}Al_{0.75}N$ |
| ***Eg*** | **0.792** | **1.029** | **4.097** |
| ***Material*** | $In_{0.96}Al_{0.04}N$ | $In_{0.86}Al_{0.14}N$ | $In_{0.25}Al_{0.75}N$ |
| ***Eg*** | **0.739** | **1.029** | **4.097** |

**Table (2)**: Bowing parameter $b_{In}$ for $In_xAl_{1-x}N$ alloys [9, 10].

| In-content (x) | $b_{In}$ (eV) |
|---|---|
| $0 \leq x \leq 0.85$ | 6 – 1 |
| $x \geq 0.25$ | 3 |

# Figures

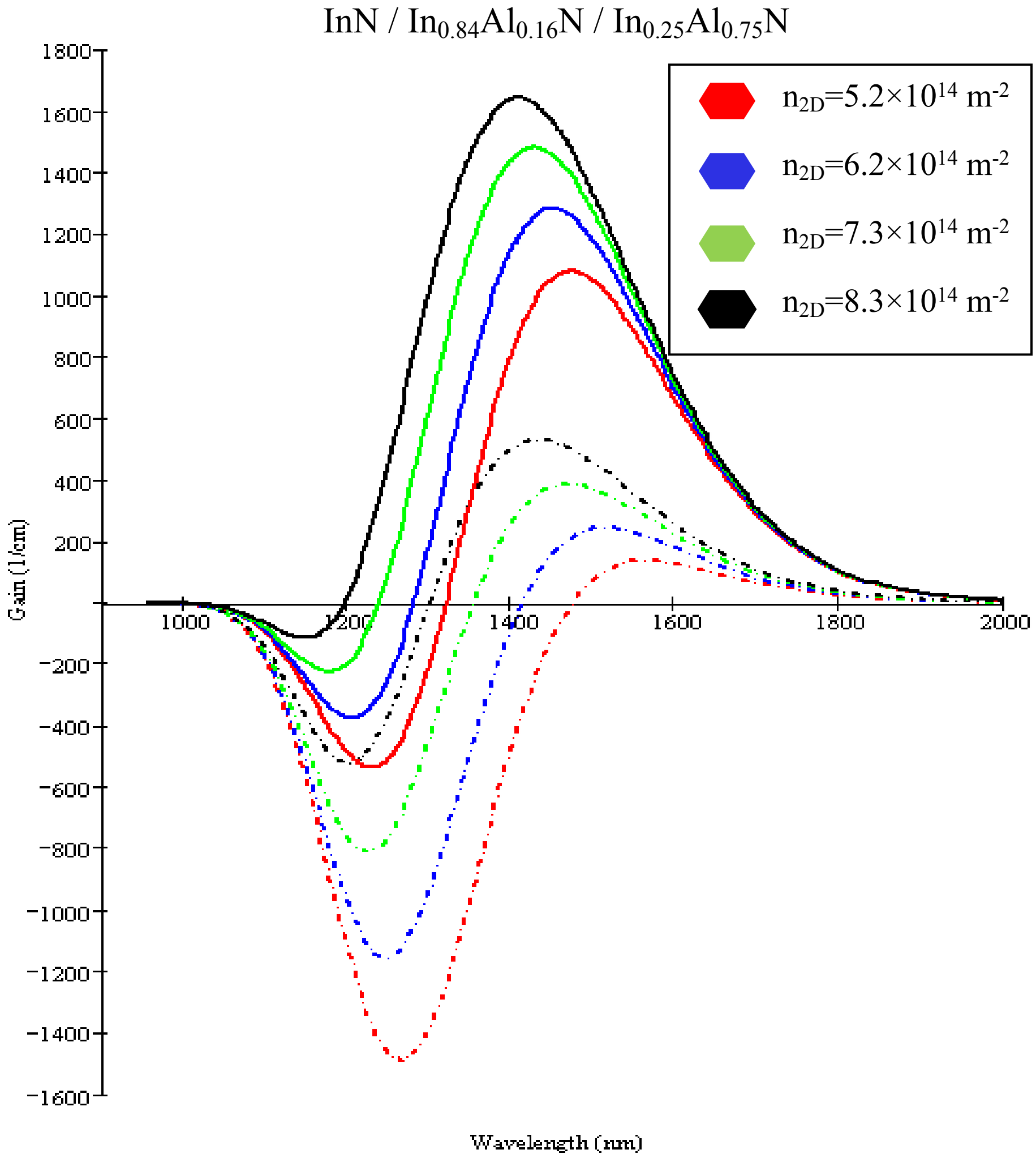

Figure 1: Calculated material gain spectra of an undoped structure (dashed lines) and p-doped structure (solid lines) for $InN/In_{0.84}Al_{0.16}N/In_{0.25}Al_{0.75}N$ structure.

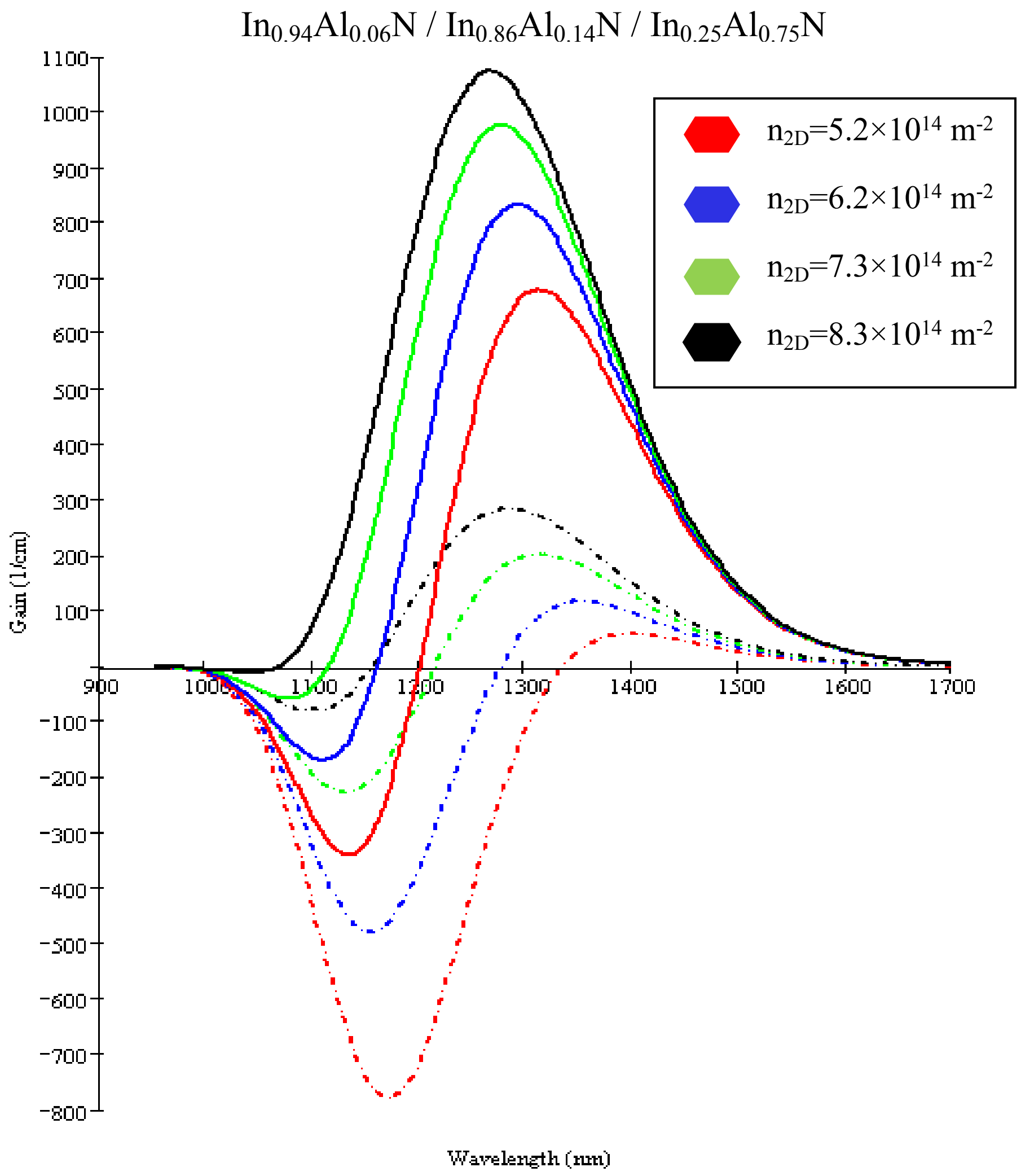

Figure 2: Calculated material gain spectra of an undoped structure (dashed lines) and p-doped structure (solid lines) for $In_{0.94}Al_{0.06}N/In_{0.86}Al_{0.14}N/In_{0.25}Al_{0.75}N$ structure.

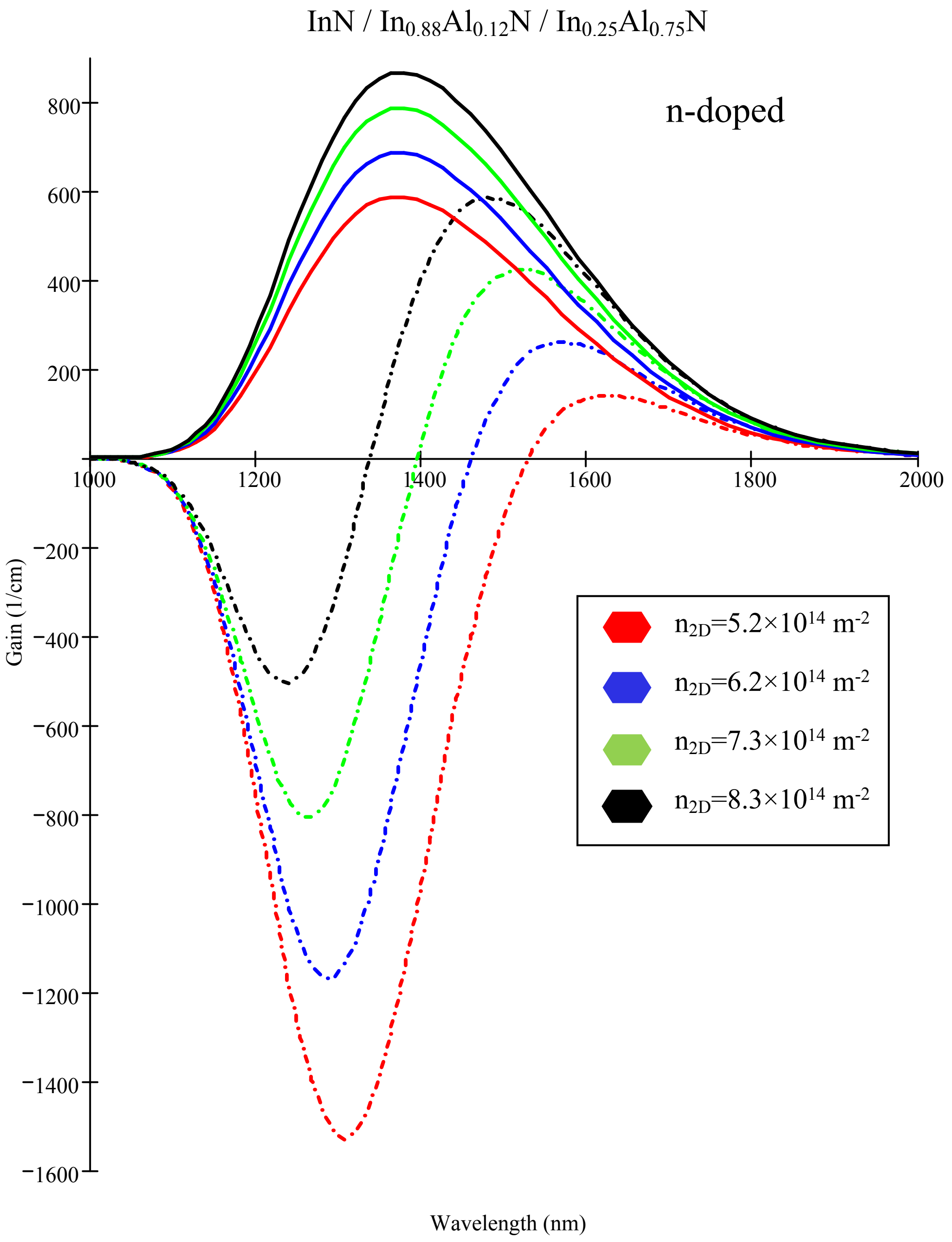

Figure 3: Calculated material gain spectra of an undoped structure (dashed lines) and an n-doped structure (solid lines) for InN / $In_{088.}Al_{0.12}N$ / $In_{0.25}Al_{0.75}N$ structure.

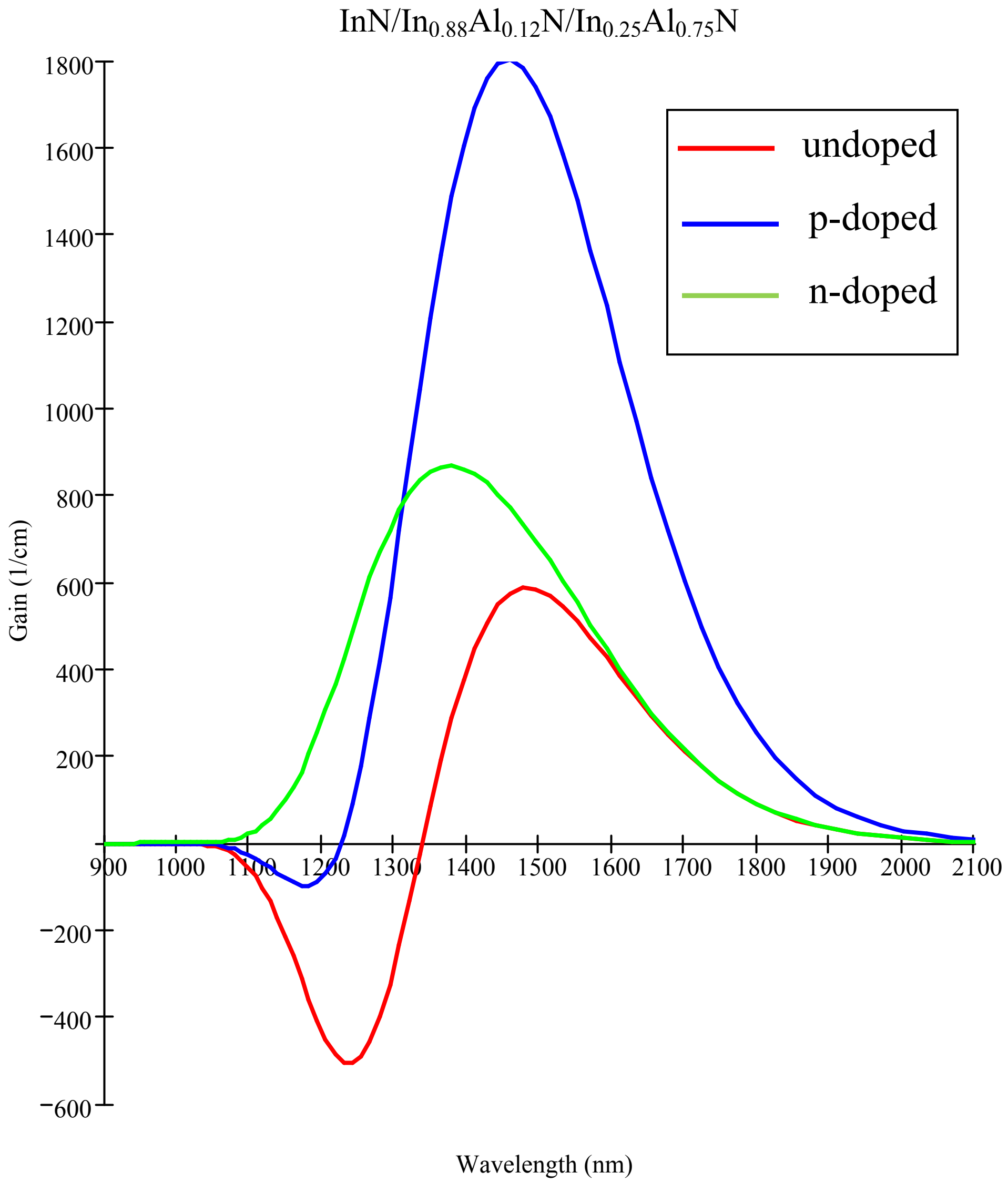

Figure 4: Calculated material gain spectra vs. wavelength compared with three material gain spectra of undoped, p-doped and n-doped, for $InN/In_{088.}Al_{0.12}N/In_{0.25}Al_{0.75}N$ structure.

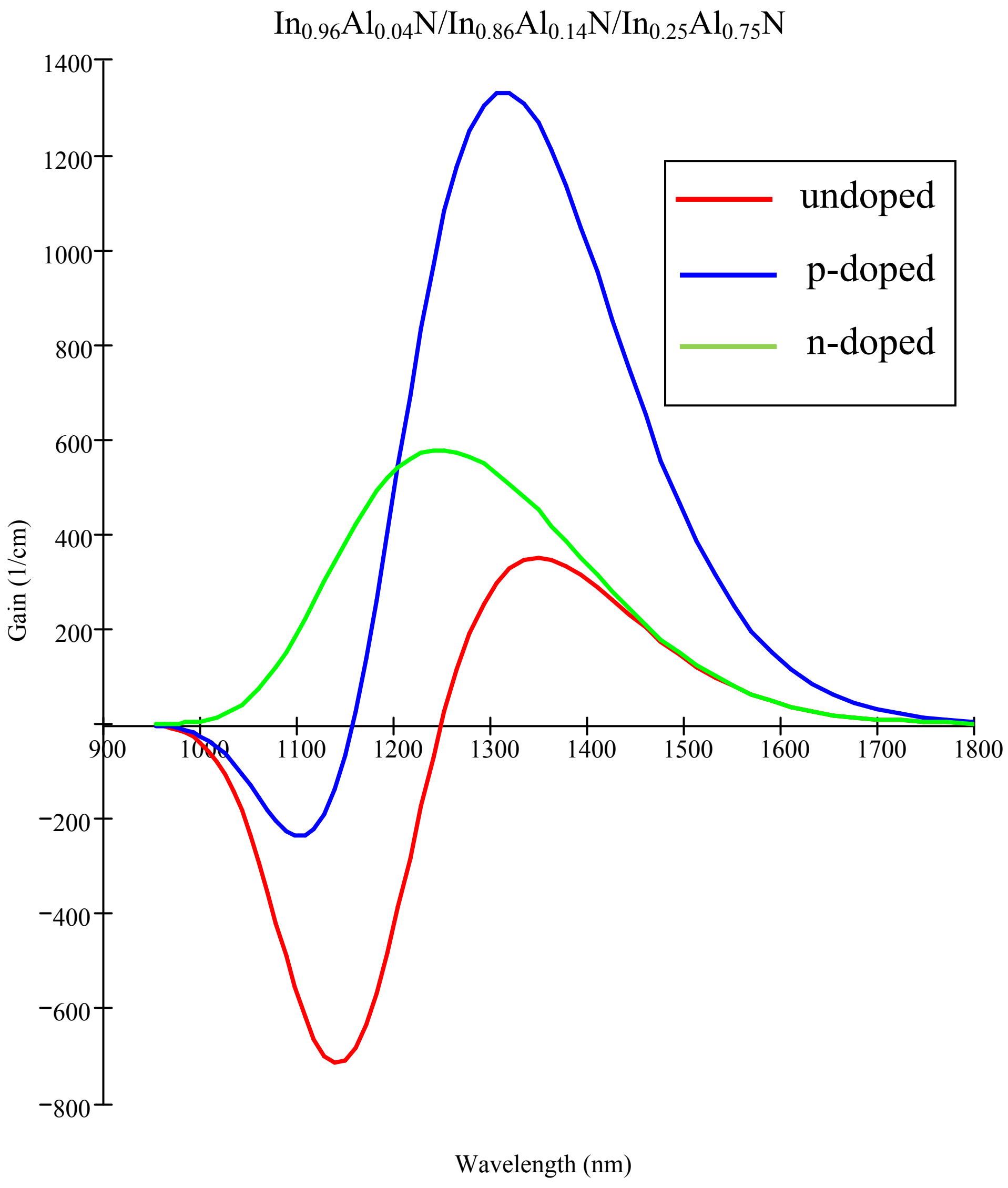

Figure 5: Calculated material gain spectra vs. wavelength for undoped, p-doped and n-doped, for $In_{0.96}Al_{0.04}N/In_{0.86}Al_{0.14}N/ In_{0.25}Al_{0.75}N$ structure.

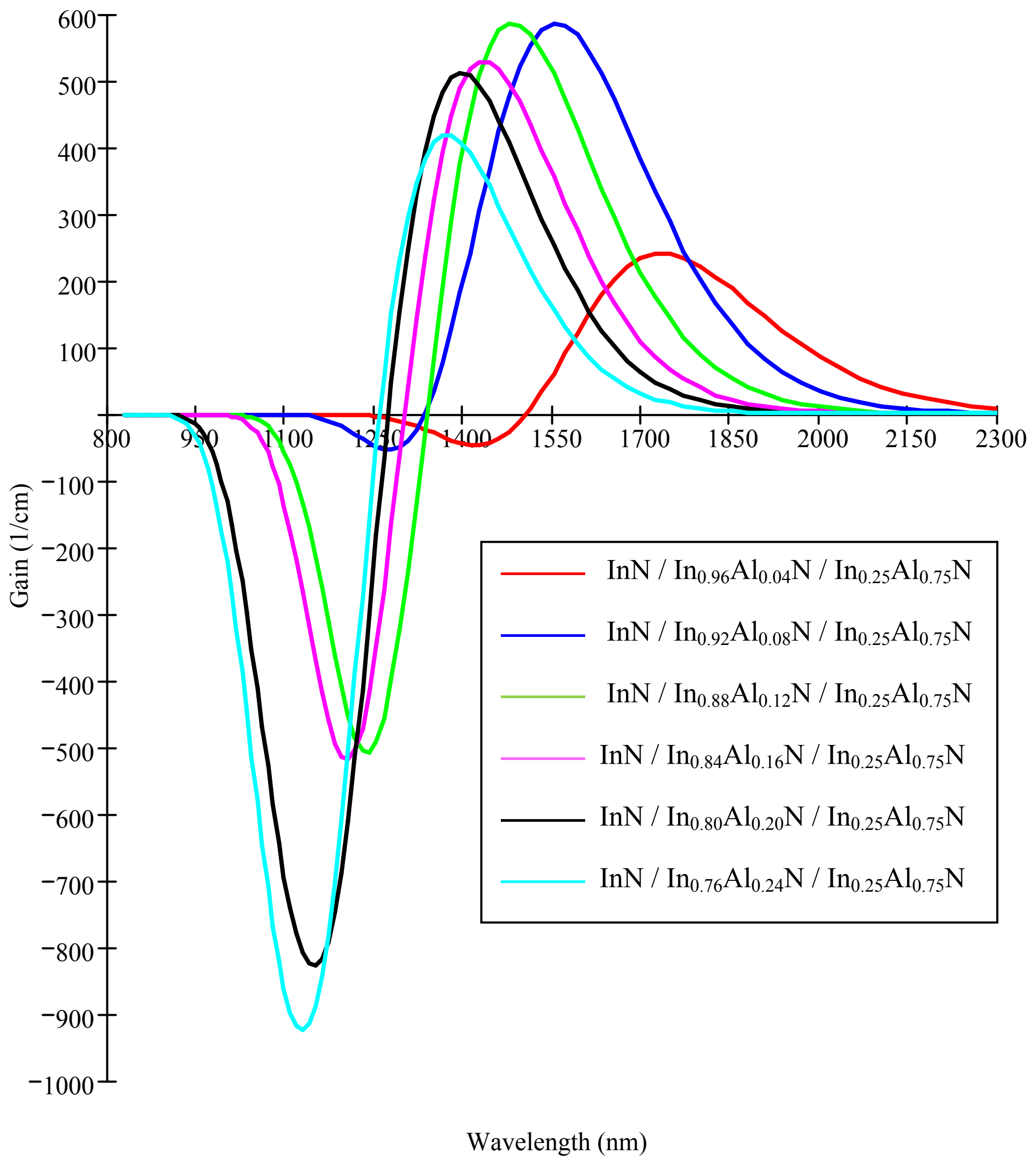

Figure 6: Calculated material gain spectra for an undoped InN QD structure at the surface electron density ($n_{2D}$=8.3×$10^{14}$ $m^{-2}$) by changing In content in the WL.

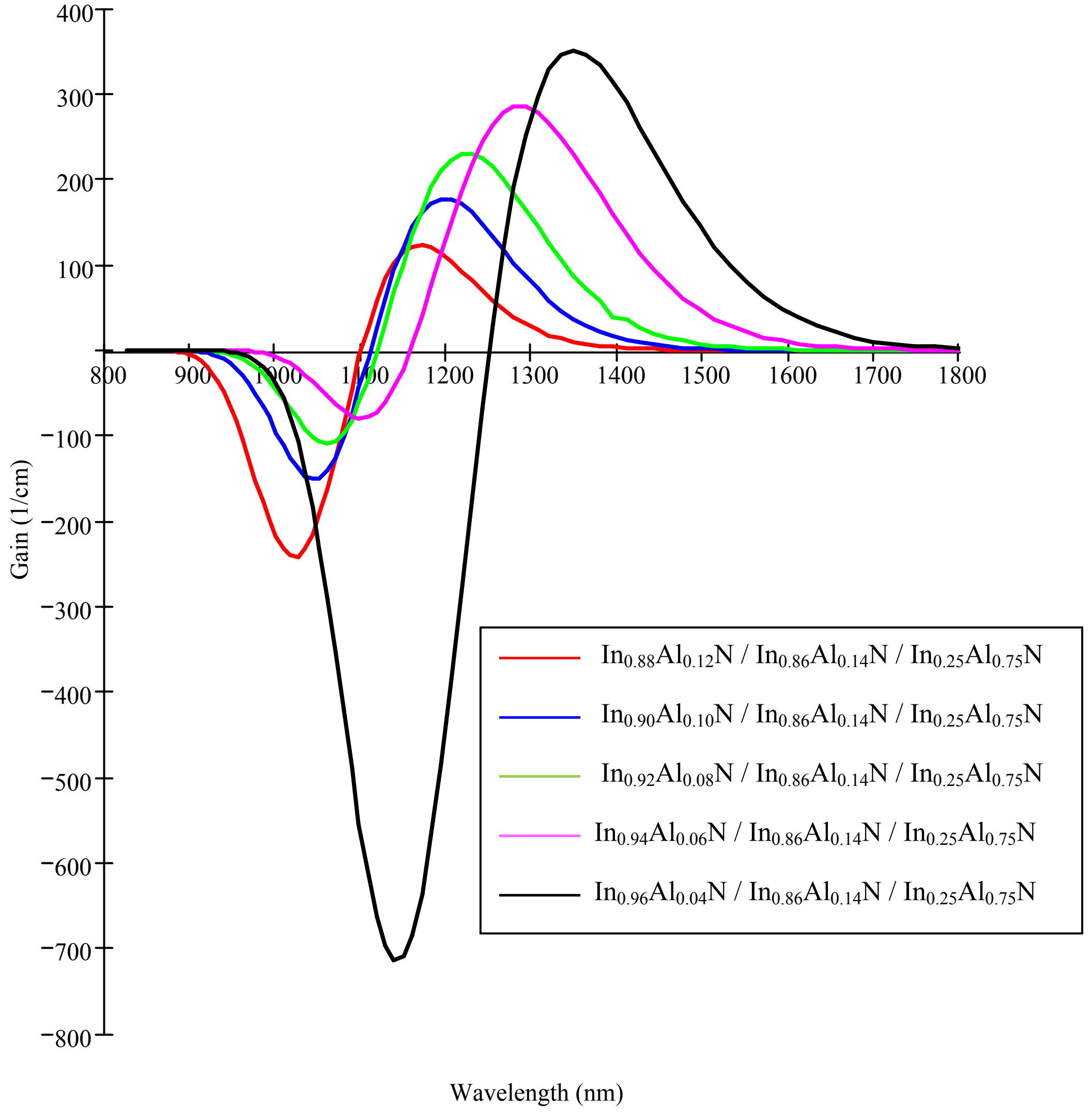

Figure 7: Calculated material gain spectra of an undoped spectra at the surface electron density ($n_{2D}$=8.3×$10^{14}$ $m^{-2}$) by changing In content in the QDs.

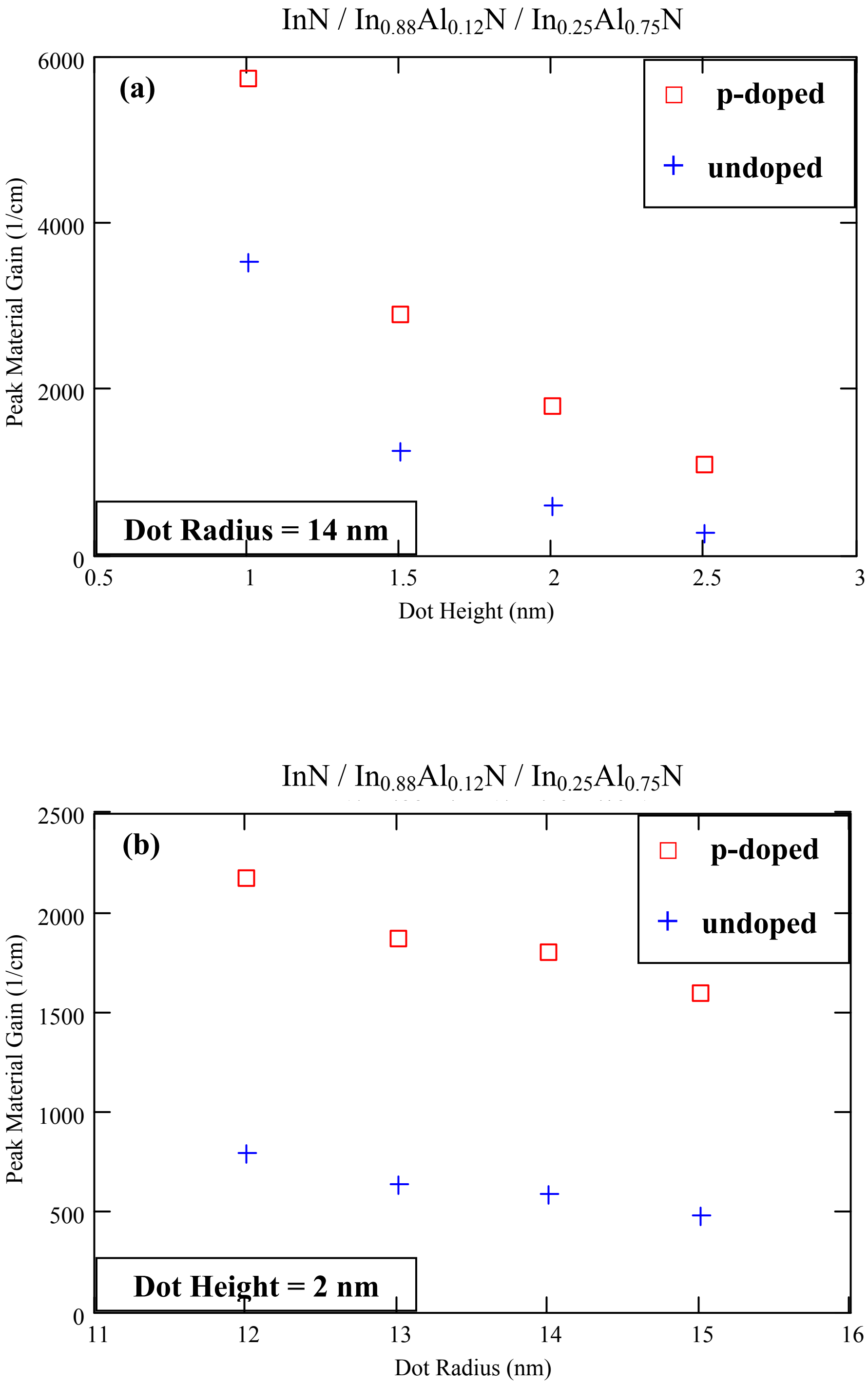

Figure 8: Calculated peak material gain vs. (a) dot height at 14 nm dot radius. (b) Dot radius at 2 nm dot height. The structures studied are undoped and p-doped InN / $In_{088.}Al_{0.12}N$ / $In_{0.25}Al_{0.75}N$.

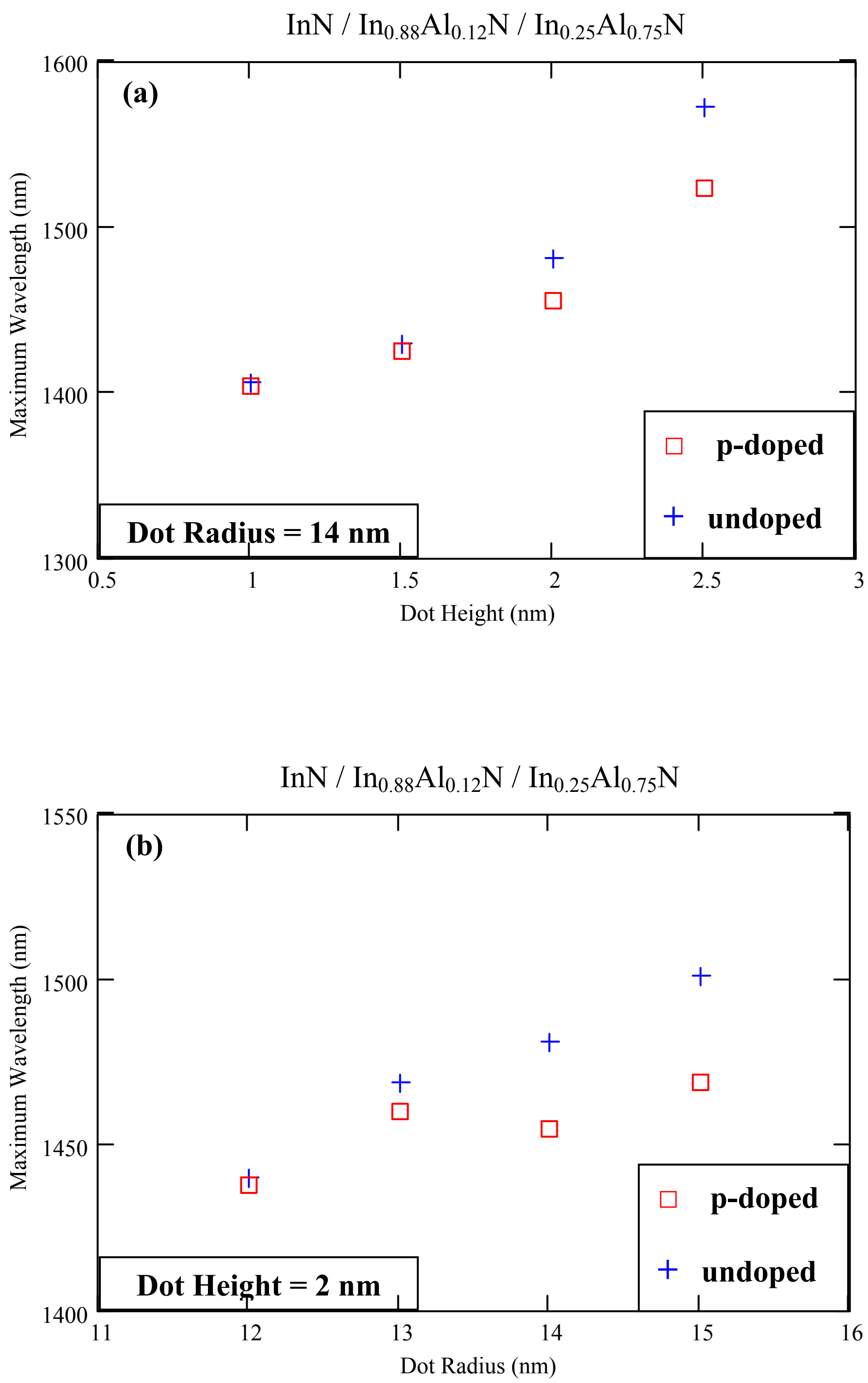

Figure 9: Calculated maximum wavelength vs. (a) dot height at 14 nm dot radius. (b) Dot radius at dot 2 nm height. The structures studied are undoped and p-doped InN / $In_{088.}Al_{0.12}N$ / $In_{0.25}Al_{0.75}N$.